\documentclass[a4paper,10pt,twoside]{cpc-hepnp}

\usepackage{multicol}
\usepackage{graphicx}
\usepackage{booktabs}
\usepackage{amssymb,bm,mathrsfs,bbm,amscd}
\usepackage[tbtags]{amsmath}
\usepackage{lastpage}

\begin{document}

\fancyhead[c]{\small Chinese Physics } \fancyfoot[C]{\small 0-\thepage}

\title{Multi- Physics analysis of the RFQ for the Injector Scheme II \\of CADS Driver Linac\thanks{Supported by National Natural Science
Foundation of China (91026001) }}

\author{%
      Wang Jing$^{1)}$\email{wangj1984@impcas.ac.cn}%
\quad Huang Jian-Long$^{1)}$
\quad HE Yuan$^{2)}$
\quad ZHANG Xiao-Qi$^{2}$\\
\quad ZHANG Zhou-Li$^{2}$
\quad Shi Ai-Min$^{2}$ }
\maketitle

\address{%
1~ Lanzhou University of Technology , Lanzhou 730000, China\\
2~{\bf }  Institute of Modern Physics, Chinese Academy of Sciences, Lanzhou 730000, China\\
}

\begin{abstract}
A 162.5 MHz, 2.1 MeV Radio Frequency Quadruples (RFQ) structure is being designed for the Injector Scheme II of China Accelerator Driver System (CADS) driver linac. The RFQ will operate at continuous wave (CW) mode as required. For the CW normal conducting machine, the  heat management will be one of the most important issues, since the temperature fluctuation may cause cavity deformation and leading to the resonant frequency  shift. Therefor a detailed multi-physics analysis is necessary to ensure that the cavity can be stably worked at the required power level.  The multi-physics analysis process includes RF Electromagnetic analysis, Thermal analysis, Mechnical analysis, and this process will be iterated for several cycles until the satisfied solution can be found. As one of the widely accepted measures, the cooling water system is used for frequency fine tunning, so the tunning capability of the cooling water system  is also studied  at different conditions. The results  indicate that with the cooling water system, both the temperature rise and the frequency shift can be controlled in an acceptable level.
\end{abstract}



\footnotetext[0]{\hspace*{-3mm}\raisebox{0.3ex}{$\scriptstyle\copyright$}2013
Chinese Physical Society and the Institute of High Energy Physics
of the Chinese Academy of Sciences and the Institute
of Modern Physics of the Chinese Academy of Sciences and IOP Publishing Ltd}%

\begin{multicols}{2}

\section{Introduction}

\begin{figure*}[!ht]
    \centering
    \includegraphics*[width=15cm]{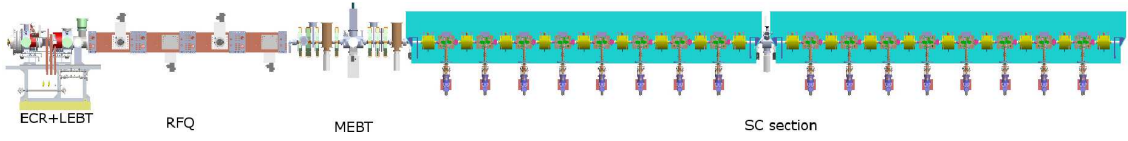}
    \caption{Layout of Injector II Linac for C-ADS.}
    \label{fig1}
\end{figure*}

Nuclear fission energy will be quickly developed  in China to both satisfy the increased requirement on energy and control the amount of the warm gas output as China energy program shows\cite{lab1}.Accelerator Driven sub-critical System (ADS) as one of the most attractive techniques to  transmute  the long-lived transuranic radionuclides produced in the reactors of the nuclear power plants into shorter-lived one, is vital for the sustainable development of the fission nuclear energy. Under such background, a China ADS project under the management of Chinese Academy of Sciences was launched at 2011, and as one of the three important parts of the China ADS, a 1.5-GeV, 10-mA continue wave (CW) proton accelerator will be developed by the Institute of Modern Physics(IMP) and Institute of High Energy Physics. The driver linac is composed by two parallel 10-MeV injectors and a main linac accelerating to the final energy.
The injector is composed of ion source, low energy beam transport line (LEBT), radio frequency quadrupole accelerator (RFQ), medium energy beam transport line (MEBT) and a superconducting accelerator section, and both the CW RFQ and the low energy superconducting structures are believed very challenge, so there are two proposals proposed and studied by two teams independently.Injector scheme II is the one that was proposed and administrated by IMP. It is characterized with a relatively low RF frequency 162.5 MHz and is composed of an ECR proton source, LEBT, 2.1 MeV RFQ, MEBT1 and superconducting section.
The layout of Injector scheme II is shown in Fig.~\ref{fig1}.
RFQ will be the key elements in the Injector II to guarantee the CW operation. The Multi-Physics analysis will be the very necessary simulation process in the RFQ design.

 The RFQ of Injector  scheme II is designed by IMP associated with the Lawrence Berkeley National Laboratory (LBNL) \cite{lab2}. It will accelerate proton beams from 35 KeV to 2.1MeV. The 162.5MHz frequency is chosen in order to decrease the power density on the surface of the cavity. The reason to apply a relative low 65 kV inter-vane voltage is also to decrease the RF dissipation power. The specifications of the RFQ are listed in Table.1.
\begin{center}
\tabcaption{ \label{tab1} The main parameters of the RFQ.}
\footnotesize
\begin{tabular*}{80mm}{@{\extracolsep{\fill}}ccccccc}
\toprule Parameters & Value \\
\hline
Ion species & Length of gap \\
RF frequency(MHz)& 162.5 \\
Inter-vane voltage V (kV) &65 \\
Total structure length (m) & 4.2079 \\
Power(kW) & 83.5 \\
Duty factor(\%) & 100 \\
\bottomrule
\end{tabular*}%
\end{center}
The structure of the RFQ is a 4-vane type. The cavity will be divided into four segments��and each of which is about 1.0 m long. The oxygen-free high-conductivity copper (OFHC) is adopted for its good performance in the heat conduction.
\section{The physics model of RFQ}

\begin{figure*}[!tb]
    \centering
    \includegraphics*[width=15cm,height=4cm]{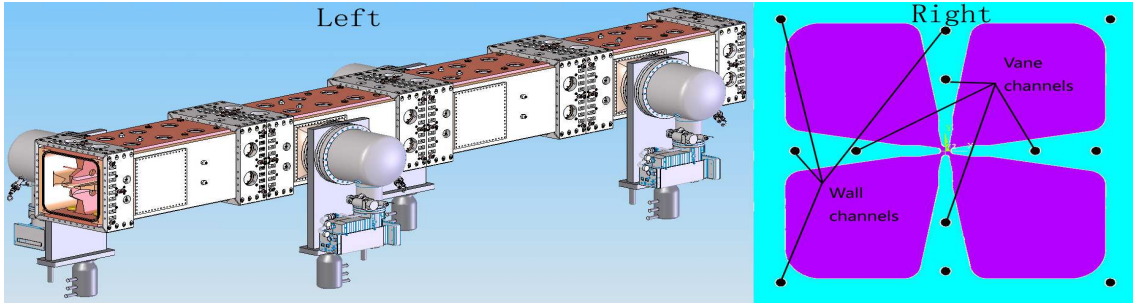}
    \caption{3-D model of the RFQ caity(Left) and cooling water channels in cross section (Rigth).There are 12 cooling water channel, 8 in the wall and 4 in the vanes.}
    \label{fig2}
\end{figure*}

The RFQ cavity will produce Joule heat by RF dissipation power during operation,  and the heat will induce the increase of the cavity temperature , if there is no efficient cooling it will lead to thermal deformation of the cavity geometry \cite{lab3}. If the thermal deformation is too large , it will lead to the detuning of the cavity. So a efficient water cooling system is necessary for the CW RFQ stable operation. For our four vane type cavity, the cooling system is consisted of  12 longitudinal coolant passages in four vanes and side and end walls. The distribution of the coolants is shown in Fig.~\ref{fig2}. The multi-physics simulations are based on this model, and the cavity is analyzed to ensure the temperature rise and frequency shift are both in an acceptable level.

\section{Multi-physics analysis procedure}
The Multi-physics analysis is an  iterative procedure including RF Electromagnetic, Thermal and mechanical analysis\cite{lab4}. In our case, the 1/4 part of the  section is  simulated considering symmetry of the structure as shown in Fig.3. The model in simulation is a 3-D slice of one quarter of the RFQ cross section, and the thickness is 1 mm in z directions. The finite element code ANSYS\cite{lab5}is used to solve such problem. The simulation procedure is described below.

\subsection{The Radio frequency analysis }

The geometry of the RF cavity is determined by the MWS\cite{lab6} code. Firstly, the RF simulation is done by the ANSYS code.  The boundary condition is electric wall and shield as shown in the Fig.~\ref{fig3}. The results of ANSYS and MWS codes are listed in Table.2 and they are agreed very well.
\begin{center}
\includegraphics[width=6cm]{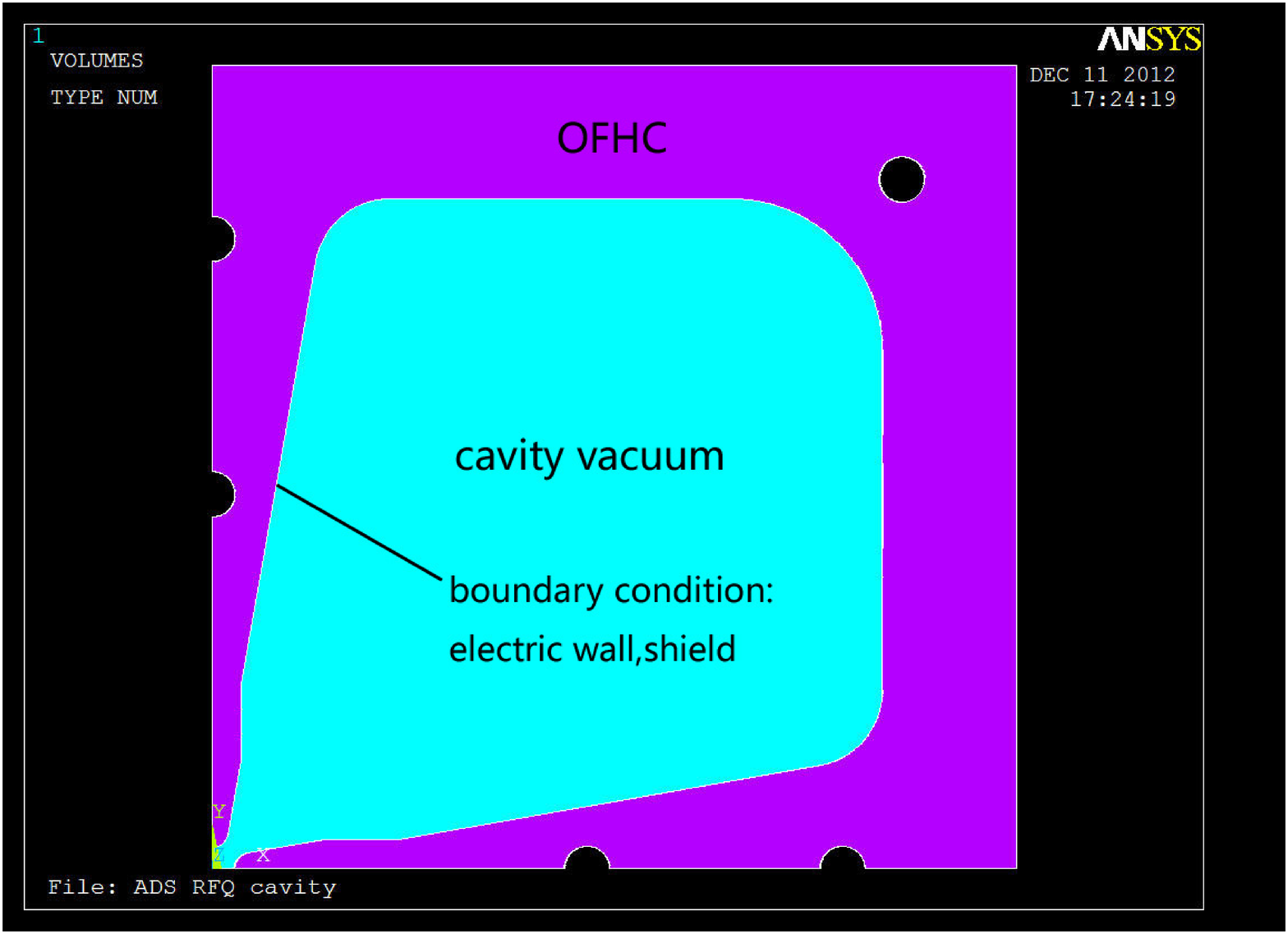}
\figcaption{\label{fig3}The RFQ model and boundary conditions of RF simulation in ANSYS. Cavity vacuum mean the part is vacuum with $\mu_0$=1 and $\epsilon_0$=1.  }
\end{center}
The results of RF simulations by ANSYS and MWS are listed in Table.2.
\begin{center}
\tabcaption{ \label{tab1} The results comparison between ANSYS and MWS.}
\footnotesize
\begin{tabular*}{80mm}{@{\extracolsep{\fill}}ccccccc}
\toprule Code & Frequency(MHz) & Q-factor\\
\hline
MWS	 & 162.569	&16861 \\
ANSYS& 162.571	&16848 \\
\bottomrule
\end{tabular*}%
\end{center}
From the comparison of the results got from both codes, the difference between the two results is around 0.1$\%$. So the heat distribution can be applied on the cavity walls for the subsequent thermal simulation.

\subsection{Thermal analysis}
The Joule heat produced by the RF power dissipation can be got from the RF analysis, which is applied to the inner cavity wall surfaces. The boundary condition is heat convection between the water and the cavity. The convection coefficient for the water cooling is evaluated prior to the simulation by the following formula\cite{lab7}:

\begin{equation}\label{1}
    h=\frac{k}{D}N {u}_{d},
\end{equation}

where $k$ is the thermal conductivity of the fluid, $D$ is the channel diameter, $N{u}_{d}$ is the Nusselt number, which is calculated using Equation (2).
\begin{equation}\label{1}
N {u}_{d}=0.023{Re}_{d}^{0.8} {P}_{r}^{0.4},
\end{equation}
${Re}_{d}$ is Reynolds number, which is the measure of relative strength of the inertial and  viscous forces in a fluid flow and can be evaluated according to  Equation (3).
\begin{equation}\label{1}
{Re}_{d}=\frac{\rho vD}{\mu },
\end{equation}
$\rho$ is the fluid density, $v$ is the fluid velocity, and $\mu$ is the dynamic fluid viscosity.And Pr is Prandlt numbers, which is related to material property.

In the simulation, the environment temperature and cooling water temperature are both set to be  20$^\circ$C and cooling water velocity is 2.29m/s. Fig.~\ref{fig4} shows the temperature profile of the calculated temperature distribution under the above mentioned conditions, and we can see the maximum temperatures is  24.9$^\circ$C and is located at the vane tips.
\begin{center}
\includegraphics[width=7.cm]{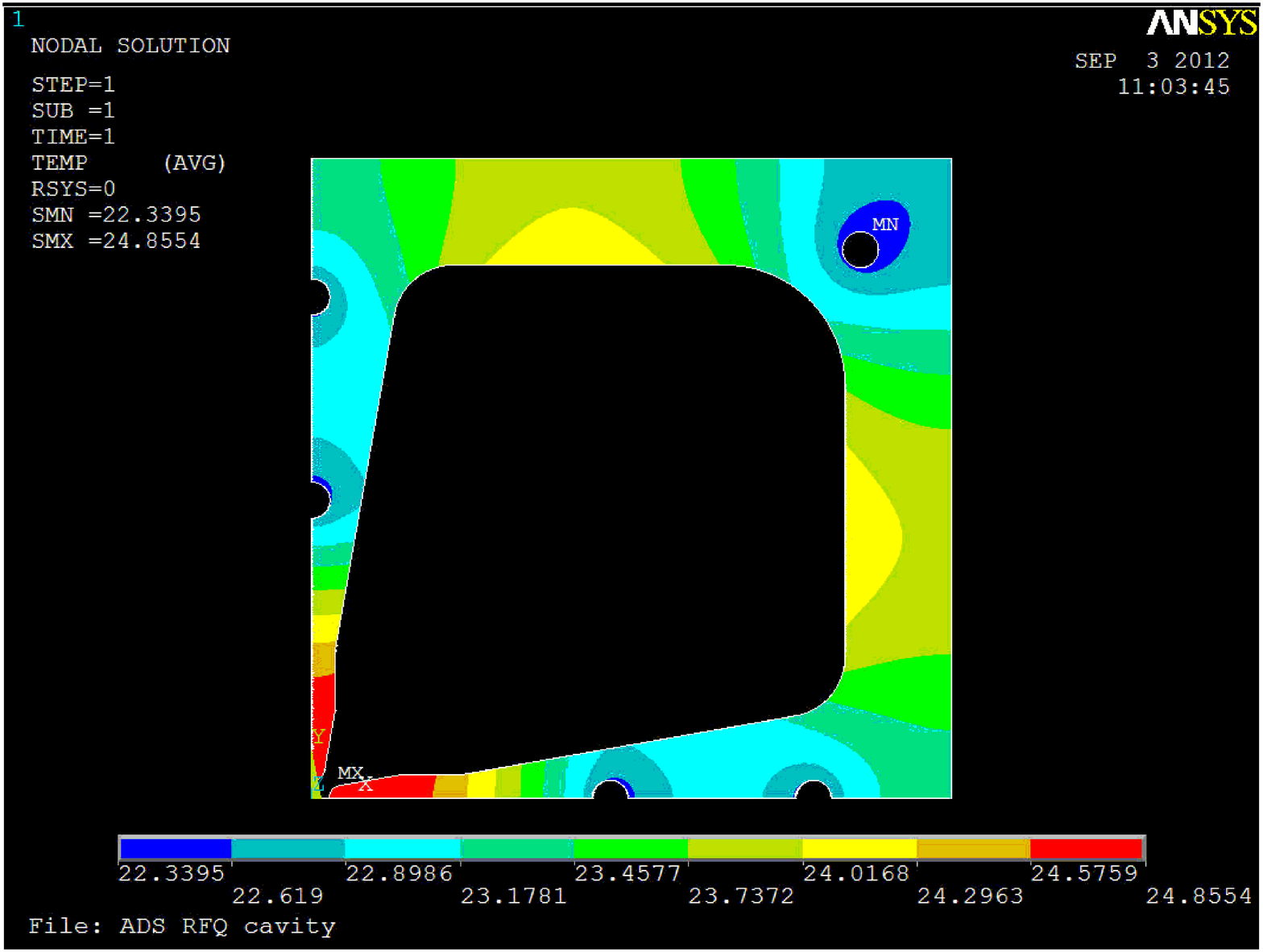}
\figcaption{\label{fig4}  RFQ Displacement Vector Sum(left) and RFQ von-Mises Stress (MPa). }
\end{center}
\subsection{Mechanical analysis}
The temperature profile simulated in step2 is used as boundary conditions for Structural analysis. ANSYS code can couple the thermal result to mechanical analysis . Nodal temperatures are applied to calculation of expansion. The mesh elements should be changed from thermal element SOLID87 to structure element SOLID187. The symmetry boundary conditions and the co-planar condition are applied to the appropriate surfaces. These conditions allow for axial growth of the RFQ cavity and accurate prediction of axial stresses.  Also, the water pressure is applied to the surface nodes of in the cooling channels and vacuum pressure is applied to the surfaces of the inner cavity walls.

Determining the displacements due to the temperature load is essential for estimating the frequency shift of the cavity when it is in operation mode. Fig.~\ref{fig5} shows the displacement vector sum due to the thermal, vacuum and water pressure loadings. Overall the displacement in the cavity is small, with a maximum of 125 ��m at the farthest corner, opposite of the vane tips. At vane tips, the displacement range is 3 to 16��m, which is quite small as well. The maximum stress in the model is calculated as 2.9 MPa, which is far below the yield strength of HOFC.

\begin{center}
\includegraphics[width=7.cm]{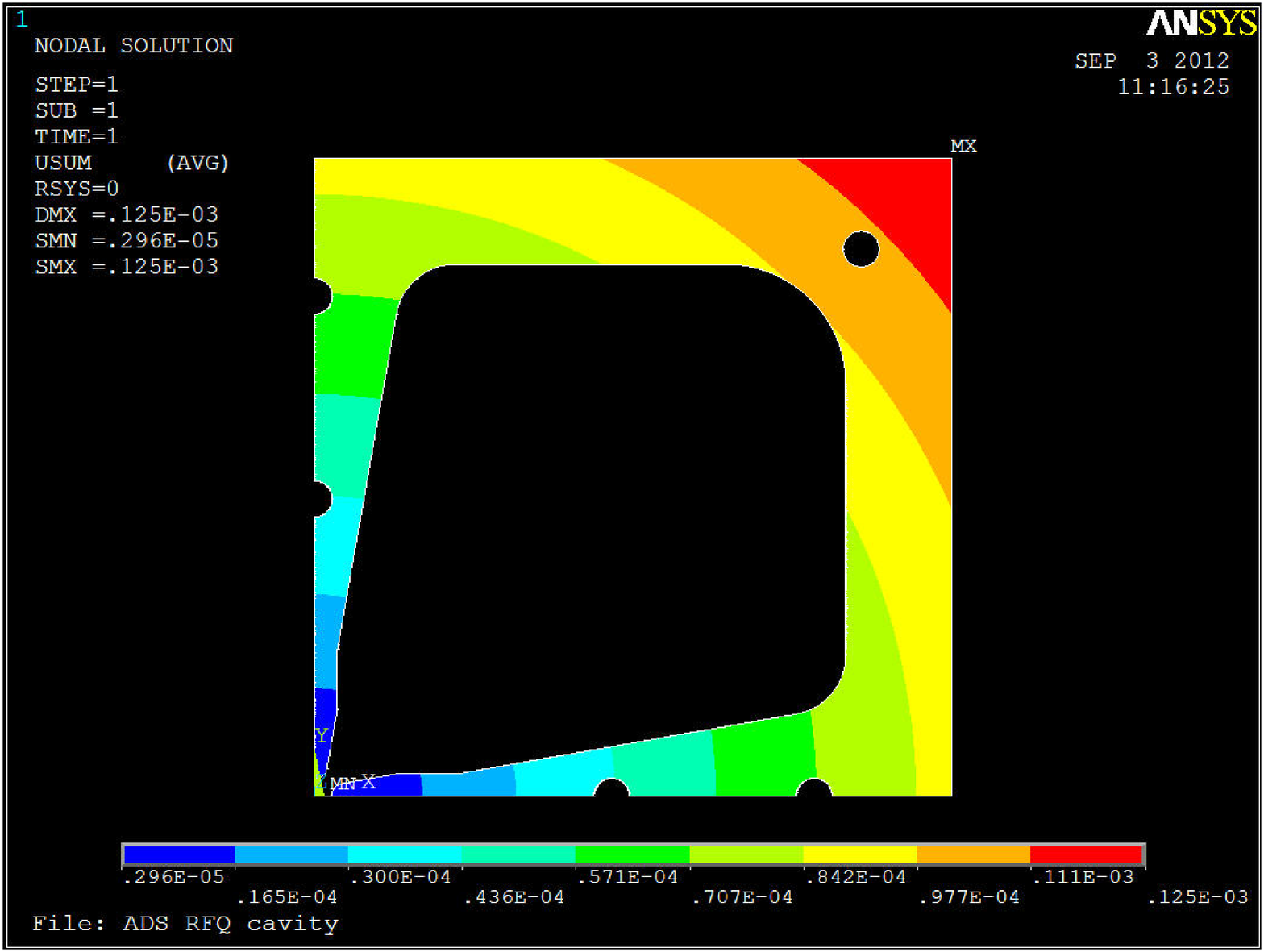}
\includegraphics[width=7.cm]{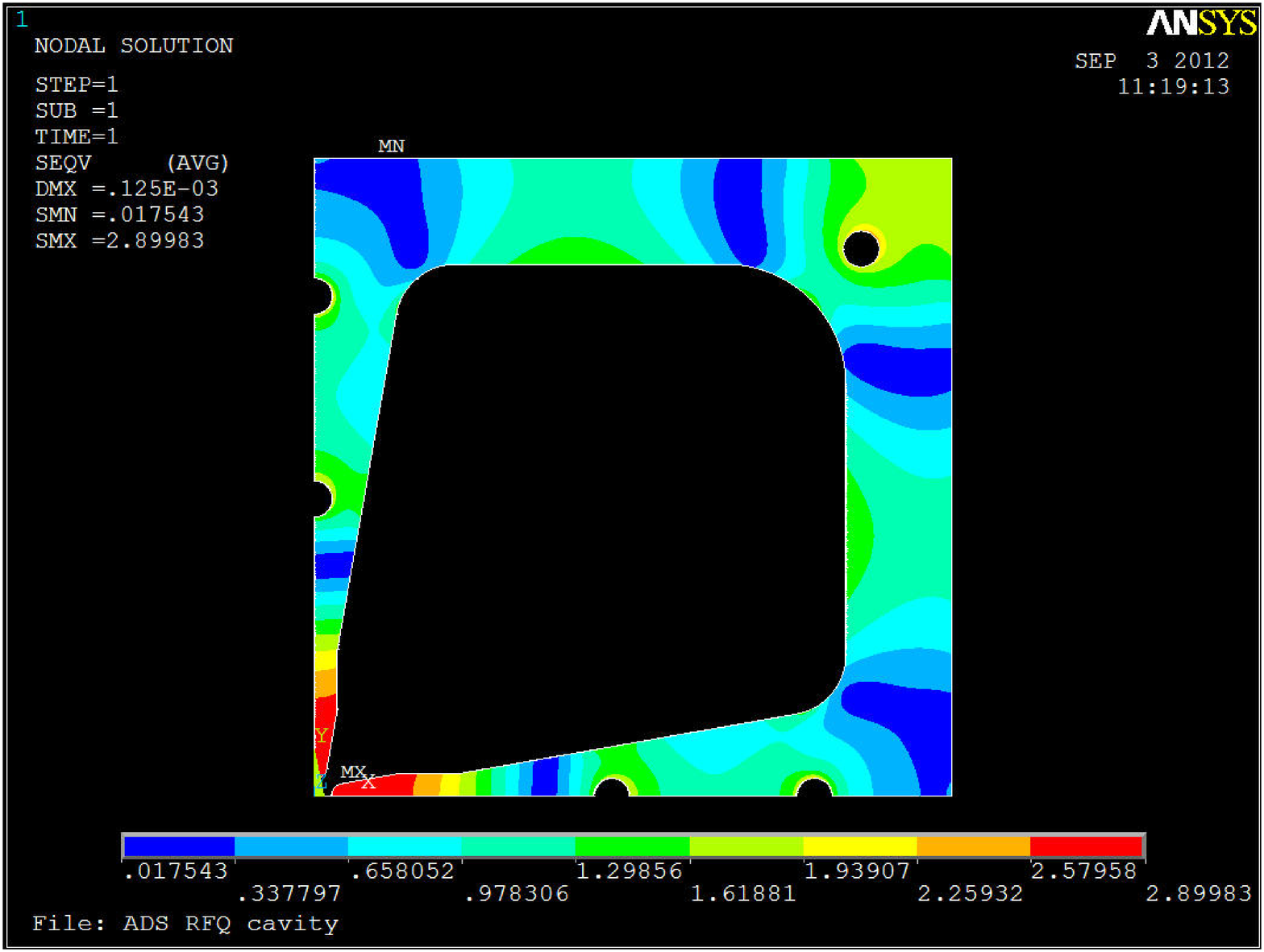}
\figcaption{\label{fig5}  Temperature profile of the RFQ simulated with ANSYS. The above picture shows the displacement distribution, and the blew one shows the stress distribution.}
\end{center}
\subsection{Final RF Analysis}
Calculating the maximum frequency shift of the RFQ depends on the deformed cavity vacuum. The final RF analysis couple the structural result back to the HF Electromagnetic analysis to determine the frequency shift of the RFQ after it has reached a steady operating condition. The deformed RF cavity geometry is used for this simulation by employing the mesh morphing method, which moves cavity nodes such that they correspond to the structural displacements, which are evaluated in the structural simulation. Compared with the calculated frequency in the first step, the frequency shift of the cavity is 108 KHz.
\section{Parameterization analysis}
In the real situation, the cool water system will not work just as the designed value, the cooling water condition and cavity status parameterization analysis is necessary. In this section, the input power of cavity, cooling water temperatures and water velocities are varied to examine how they affect heat dissipation and result in frequency shift of the cavity. This is useful as it can help to determine the RFQ��s sensitivity to these actual conditions and can give us a reference in how to adjust the cooling water temperature when the RFQ cavity is under RF power training and beam commissioning. The results under different conditions are illustrated in the following.
The base boundary condition: environment temperature and cooling water temperature are set to be  20$^\circ$C,cooling water velocity is 2.29m/s. There is one note that the frequency shift is set to be zero when the cavity at the base boundary condition.
\section{Parametrization analysis}
\subsection{Water velocity}
The cooling water velocity is changed based on 2.29m/s to study its effect on frequency shift. In our water cooling system, the water channel in vane and wall are controlled separately. The effect of wall and vane cooling water on frequency drift are plotted in the Fig.~\ref{fig6}. From the figure, when the water velocity changes from 2m/s to 2.7m/s, the frequency shift is smaller than 6.0 KHz, which is below the band width of the RFQ cavity.
\begin{center}
\includegraphics[width=7.cm]{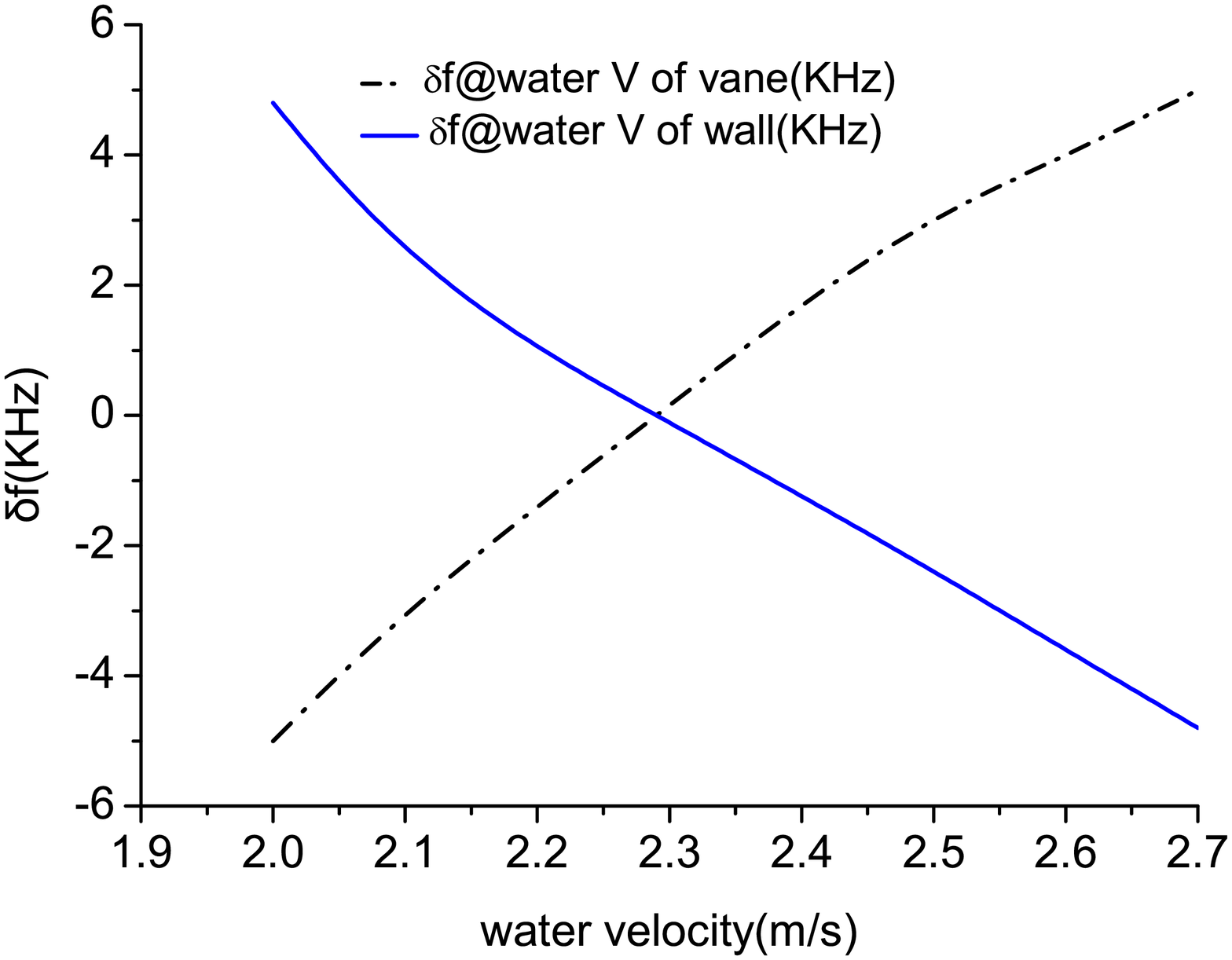}
\figcaption{\label{fig6}Relations between cavity frequency and water velocity, the solid line is the frequency shift from wall water velocity, the dash dot line from vane water velocity. }
\end{center}
\subsection{Input RF power}
In the analysis, the cavity temperature increase, maximum displacement, maximum stress and frequency shift caused by input RF power are studied. In the Fig.~\ref{fig7}, the cavity temperature increase, maximum displacement, maximum stress all vary linearly with input RF power, while the frequency shift case is inverse. This is because the cavity dimension will enlarge slightly which will induce the cavity frequency decrease. The frequency shift is less than 10 KHz between cold and full power cavity. The simulation can also give us a reference to reduce reflect power by change the frequency of power amplifier.
\begin{center}
\includegraphics[width=7.5cm]{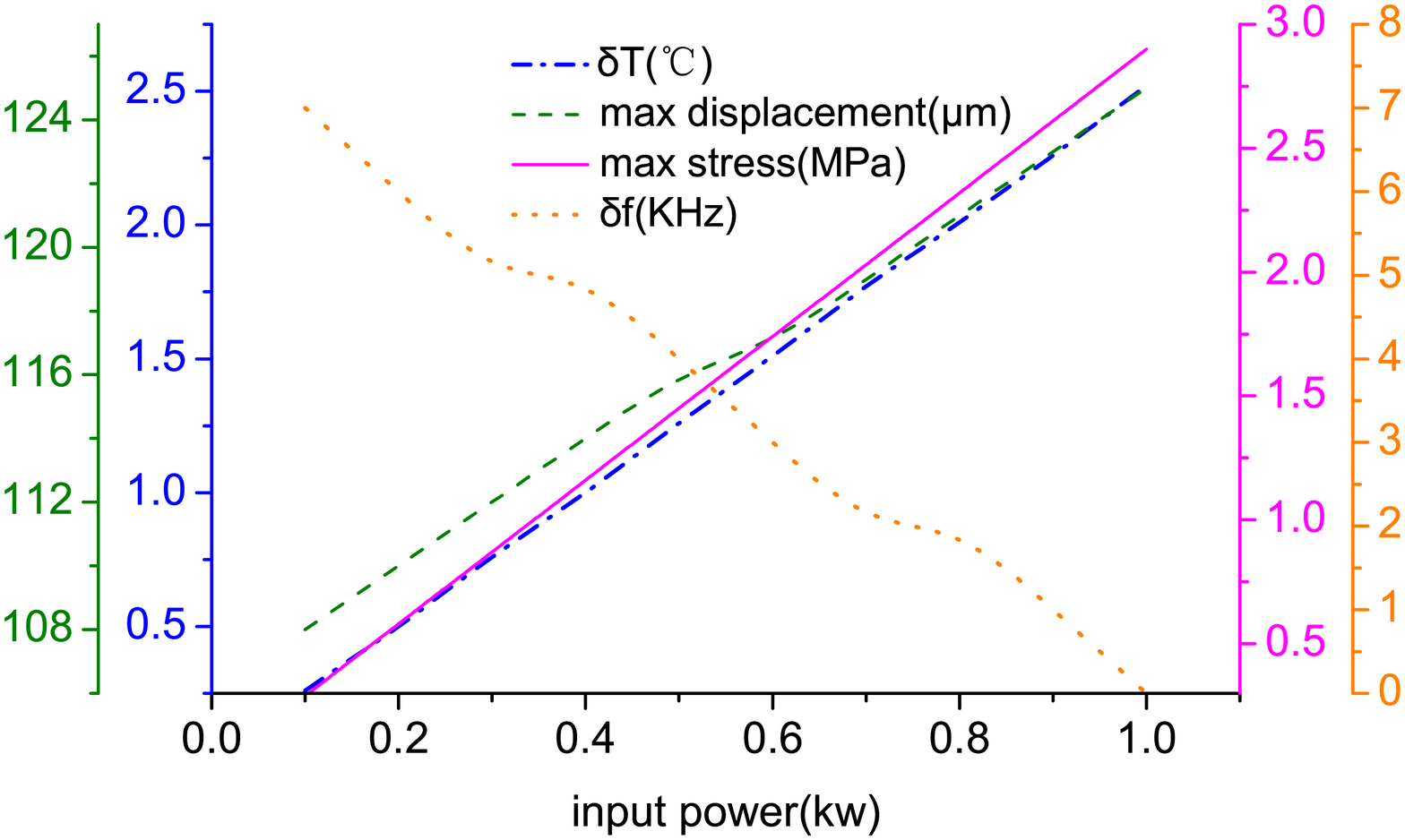}
\figcaption{\label{fig7}Relationships between input power and temperature rise, displacement and stress, and frequency shift. The dash dot line is temperature rise, the dash line is maximum displacement, the solid line is maximum von-mises stress, the dot line is frequency shift by changing the input power. }
\end{center}
\subsection{Cooling water temperature}
The effects of cooling water temperature from vane, wall and total are separately shown in Fig.~\ref{fig8} with different curves. From the figure, the frequency will decrease at the ratio -16.125KHz/$^\circ$C as the temperature of cooling water in the vane. While  the effect from cooling water in the wall will be inverse at the ratio 12.875KHz/$^\circ$C.The frequency drift from the total cooling water change will be goes down at -3KHz/$^\circ$C. Form the simulation, the frequency drift is more sensitive to the cooling water in the vane. This gives us a guide that the wall water temperature will be fixed to adjust the vane water to tune the cavity frequency.
\begin{center}
\includegraphics[width=7.cm]{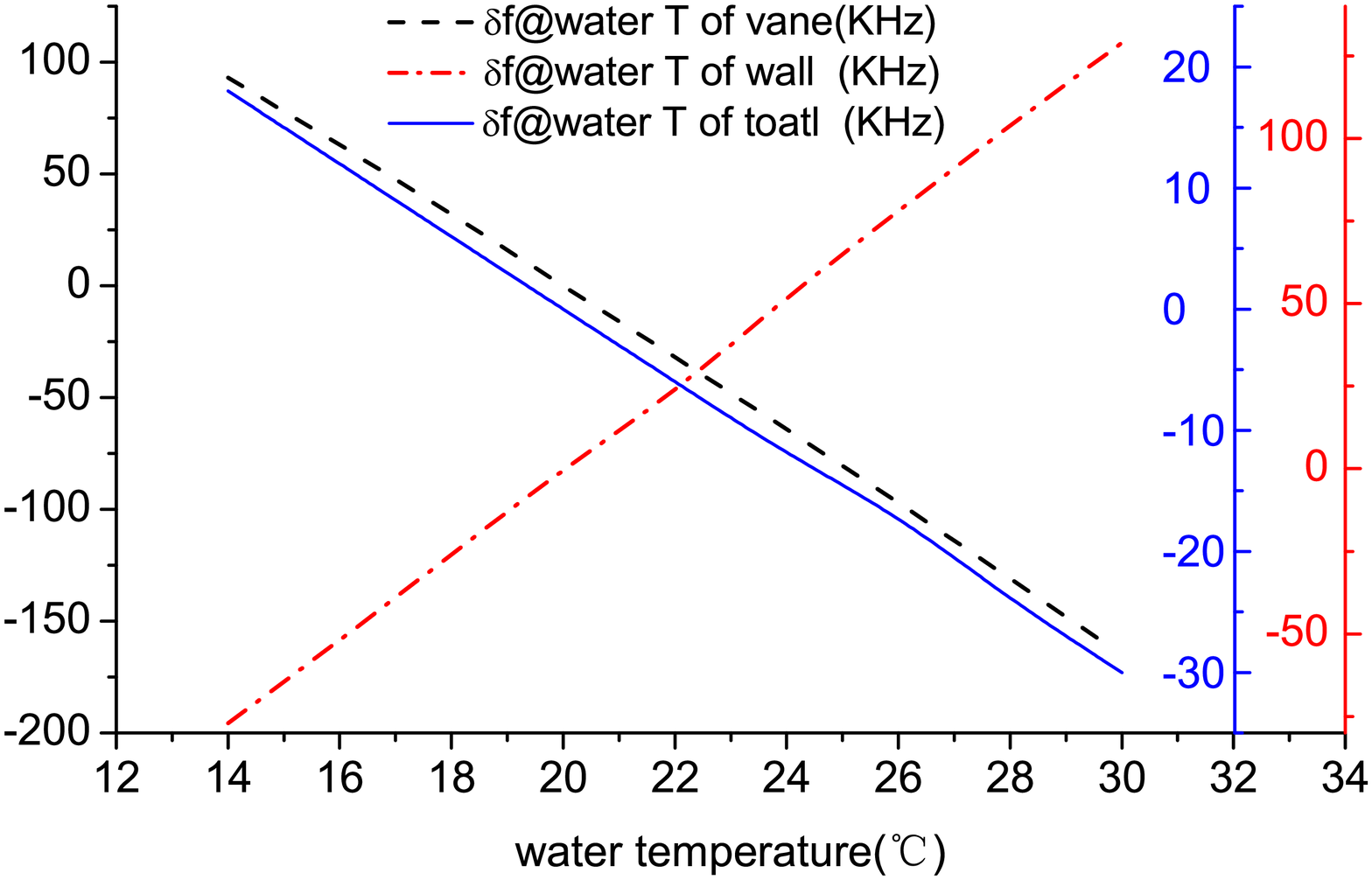}
\figcaption{\label{fig8}Frequency shift responding to water temperature. The dash line is the frequency shift from vane water temperature, the dash dot line is the frequency shift from wall water temperature, the solid line is the frequency shift from total water temperature. }
\end{center}
\section{Summary and outlook}
The processes of multi-physics analysis are presented in the paper. The analysis processes can also be used in the other room temperature RF structures. The parameterization analysis of input power, cooling water temperature and velocity are studied. The results show that the cooling water system can meet the requirements of RFQ cavity operated at full power. Also from the parameterization analysis, we can get reference of tuning the cavity at RF power training and beam commissioning.
	
The 2-D simulation has been done, more accurate 3-D model simulation will be simulated and studied in the further work.

\acknowledgments
{One of the authors Jing Wang would express his sincere thanks to Dr Andrew Lambert(LBNL) for his useful talk and suggestion on the use of ANSYS code.}

\end{multicols}
\vspace{10mm}

\centerline{\rule{80mm}{0.1pt}}
\vspace{2mm}
\begin{multicols}{2}

\end{multicols}

\clearpage

\end{document}